\documentclass{llncs}
\usepackage{amsmath}
\usepackage{myref}
\usepackage{graphicx}
\usepackage{subfigure}
\usepackage{algorithm}
\usepackage{algorithmic}
\usepackage{color}
\usepackage{listings}
\usepackage{multirow}
\usepackage{rotating}
\usepackage{verbatim}
\usepackage{makeidx}  
\begin{document}

\setlength{\pdfpagewidth}{8.5in}
\setlength{\pdfpageheight}{11in}

\renewcommand\t[1]{{\tt #1}}
\mainmatter              

\title{Fault Localization Using Textual Similarities}

\author{Zachary P. Fry \and Westley Weimer \\
\email{\{zpf5a,weimer\}@cs.virginia.edu} 
\institute{University of Virginia}
}

\maketitle


\begin{abstract}

Maintenance is a dominant component of software cost, and localizing
reported defects is a significant component of maintenance. We propose a
scalable approach that leverages the natural language present in both
defect reports and source code to identify files that are
potentially related to the defect in question. Our technique is
language-independent and does not require test cases. The approach
represents reports and code as separate structured documents
and ranks source files based on a document similarity metric
that leverages inter-document relationships.

~~~We evaluate the fault-localization accuracy of our method against both
lightweight baseline techniques and also reported results from
state-of-the-art tools. In an empirical evaluation of 5345 historical
defects from programs totaling 6.5 million lines of code, our approach
reduced the number of files inspected per defect by over 91\%.
Additionally, we qualitatively and quantitatively examine the utility of
the textual and surface features used by our approach. 


\end{abstract}

\section{Introduction}
\label{sec-intro}

Maintenance tasks can account for up to 90\% of the overall cost of
software projects~\cite{boehm,Erlikh}. A significant portion of that cost
is incurred while dealing with software
defects~\cite{ramamoothy}. Large software projects typically use defect
reporting systems that allow users to submit reports directly; this
has been shown to improve overall software
quality~\cite{anvik05,Raymond02}. User-submitted defect reports vary
widely in utility~\cite{bugquality}; reports go through \emph{triage}
to allow developers to focus on those reports that are most likely to
lead to a resolution. We propose a system to make the maintenance
process more efficient by reducing the cost of localizing faults by leveraging
this user-provided information. 

\emph{Fault localization} is the process of mapping a fault (i.e.,
observed erroneous behavior) back to the code that may have caused
it. Performing fault localization is relatively time
consuming~\cite{vessey} and thus costly. For this reason, many
existing techniques attempt to facilitate this process. In general,
such techniques rely on test
cases~\cite{Agrawal95,cleve05,harrold05,Renieris03,wang09}, model
checking~\cite{ball03symptom,ball02}, or remote
monitoring~\cite{Liblit:2003:BIRPS,liblit05}. These approaches may not
be directly applicable to user-submitted defect reports, since reports
rarely provide a full test case or program trace~\cite{bugquality}.

In this paper, we address the cost of such localization for
user-submitted defect reports. We present a lightweight approach that
maps defect reports to source code locations. Our approach relies
primarily on textual features of both source code and defect report
descriptions, although it takes advantage of certain additional
information (e.g., stack traces, version control histories) when 
they are available. Notably, it does not require test cases,
compilation, execution
traces, or remote sampling, all of which can potentially limit the
applicability of other fault localization strategies.\footnote{Strictly
speaking, this paper presents a \emph{defect} localization approach, 
but \emph{fault localizaton} is the term of art used for this line of
research (e.g.,~\cite{Agrawal95,harrold05,Renieris03,wang09}).}

Our approach is based on several underlying assumptions about the
textual features of both source code and defect reports. With respect to
code, we assume that developers choose identifier names and comment
text that
are representative of observable program behavior. For defect reports,
we assume that the reporters use a vocabulary based on their
observations of program behavior --- a vocabulary that will thus be
in some ways similar to developers', although 
reporters may not have access to the source. Finally, we hypothesize
that a defect report and a code location are more likely to pertain to
the same fault if they are similar in terms of word usage; we
formalize this in a similarity metric.  

The main contributions of this research project are thus: 
\begin{itemize}

\item A lightweight, language-independent model that statically
measures similarity between defect reports and source files for the
purpose of locating faults. This comparison is based on a structured textual
analysis of the natural language in both documents.

\item A large empirical evaluation of our technique 
  including 5345 real-world defects from
  three large programs totaling 6.5 million lines of code --- over an
  order of magnitude larger than the evaluations in previous
  work~\cite{cleve05,harrold05,Renieris03}. 
  Our approach reduces
  the number of files developers inspected per defect by 91.5\%,
  outperforming baselines such as using user-reported stack traces
  (53.1\%) and previously-published results. 

\item A quantitative and qualitative explanation of our technique's
  success. Notably, we show that factors such as the reported priority
  or the number of duplicate reports present are not related to our
  model's success. Instead, we find that human word choice determines
  the performance of our model, thus supporting our hypothesis that the
  vocabulary chosen by developers and reporters can be used to localize
  faults. 
\end{itemize}

The structure of this document is as follows. In \secref{motivation}, we
motivate our approach by presenting an example fault with its
associated defect report and source code. \secref{model} outlines our
approach and formally defines how we measure the relative similarity
between code and defect report text. Next, \secref{eval} presents a detailed 
empirical 
evaluation of our approach. \secref{related} 
places our work in context. 
Finally, \secref{future} concludes.

		
\section{Motivating Example}
\label{sec-motivation}
\label{sec-example}

In this section, we present an example defect report taken from the
Eclipse project. This example illustrates the potential benefit of
matching the natural language in a defect description with keywords from
the source code for the purpose of identifying the defect's location. 

User-submitted defect reports typically consist of a free-form textual
description of the fault. When presented with such a defect report, it is
up to the developer to derive and locate the cause of the undesirable
behavior. This requires thorough familiarity with
the code base; for large projects, an important part of the triage
process is finding which developer is most likely to be able to
resolve a given defect report~\cite{anvik06}. There are significant
differences among developers in terms of how quickly they can locate a
given fault~\cite{vessey}. Our goal is to narrow the
source code search space that the developer needs to consider, 
thereby
decreasing the software maintenance cost overall. 

Consider the following defect report from the Eclipse project, defect
\#91543, entitled ``Exception when placing a breakpoint (double click on
ruler).''  The description is as follows:

{
\footnotesize
\begin{verbatim}
    With M6 and also with build I20050414-1107 
    i get the stacktrace below now and then when
    wanting the place a breakpoint when double
    clicking in the editor bar.  if i close the 
    editor and reopen it again it goes ok.

    !MESSAGE Error within Debug UI: 
    !STACK 0
    org.eclipse.jface.text.BadLocationException
          at
    org.eclipse.jface.text.AbstractLineTracker.get-
    LineInformation(AbstractLineTracker.java:251)
    ...
\end{verbatim}
}

Initially, a developer might be inclined to inspect code implicated
directly.  In this case, one might check the
\emph{AbstractLineTracker} file and other files in the stack trace, or
search the list of all files that reference a
\emph{BadLocationException}. Additionally, one might scan the files
that were changed prior to either of the particular builds mentioned.
Finally, based on basic searching (using a tool such as \emph{grep}) one 
might uncover any of the
following files: \emph{Breakpoint.java},
\emph{MethodBreakpointTypeChange.java},
\emph{BreakpointsLocation.java}, and \emph{TaskRulerAction.java}, among
hundreds of others. This example illustrates that
the search space is large, even when a programmer uses the defect 
report's specific information.

In the actual patch for this defect, developers edited only two source files.
\emph{ToggleBreakpointAction.java} contained the majority
of changes that addressed this defect report, with one
minor change to a call-site in
\emph{RulerToggleBreakpointActionDelegate.java}.  Some of the methods
in those files include: 


{ \footnotesize
\begin{verbatim}
  ToggleBreakpointAction(..., IVerticalRulerInfo
         rulerInfo)
  ToggleBreakpointAction.reportException(
         Exception e)
  RulerToggleBreakpointActionDelegatecreateAction(
         ITextEditor editor, IVerticalRulerInfo 
         rulerInfo)
\end{verbatim} 
}

The identifier names associated with these two files show clear
language overlap with the report above. For example, even when only
the report title and the method names are considered, key words
such as \emph{breakpoint}, \emph{exception} and \emph{ruler} occur
in both sets. When examining the overall word similarity,
the two files that were
changed for the fix are among those files most similar to
the text in the defect report. Using textual similarity not only
avoids unrelated methods considered by traditional search
techniques~\cite{hill09}, but
further limits the fault localization search space by trimming files
with coincidental or narrow language overlap. Aggregating overall word
similarity ensures that only documents with considerable and
meaningful similarity are favored. 

The log messages associated with software repositories represent
another possible source of human-chosen natural language information
to leverage when localizing faults. The
\emph{ToggleBreakpointAction.java} file accumulated seventeen log
messages over four years worth of changes.  Examples of these log
messages include:
\begin{itemize}
\item \t{"Can't set a breakpoint on the first line of an editor"}
\item \t{"Allow multiple debuggers to create breakpoints using the} \\ \t{same editor."}
\item \t{"NullPointerException when trying to set} \\ \t{breakpoint (in 
ToggleBreakpointAction)"}
\end{itemize}

Terms such as \emph{breakpoint}, \emph{editor} and \emph{exception}
occur in both the log messages and the defect report, suggesting that
this file may be relevant.  Repository log messages are typically
written in plain natural language which can be extracted with minimal
analysis effort, much like comments in source code. Furthermore,
repository log messages are written to chronicle the changes made to a
given source code file. In addition, as many of these changes attempt
to address previous defects, it is reasonable to assume they might
contain specific terms taken from previous defect reports themselves. 

We hypothesize that prioritizing the search space by ranking files of
interest in this manner can greatly facilitate fault localization ---
using only static, natural language information, such as the defect
report, source code and log messages. In the next section, we present
a model to take advantage of this intuition.


\section{Methodology}
\label{sec-model}
Our goal is to reduce the cost of software maintenance, focusing
on fault or defect localization. 
The available input includes a defect report describing a fault, as well
as static textual software development artifacts, such as the project
source code and revision history. The desired output is an 
ordered list of source files that are likely to contain
the cause of that fault.  

To fix the defect in question, such a list can be explored directly or
further refined, depending on size of the system and resources available.
While the resulting lists can still be quite large and must be processed
manually, previous work has shown that such filtering localizations are
helpful~\cite{ball03symptom}. More specifically, over a similar set of
defect reports accompanied by lists of methods (i.e., backtraces or
counterexamples), humans were shown to take less time to address those
defect reports in which an additional tool-generated annotation narrowed
that information down to a smaller number of lines~\cite[Fig.~5]{weimer06}.
Fault localization information might also be used as an attachment on the
original defect report: in a study of over 27,000 historical defect
reports, those including similar attachments and comments (e.g., backtraces
or lists of methods) were more likely to be resolved
rapidly~\cite[Fig.~7]{bugquality}. 

Since defect reports and development text have different formats, we
propose to map both of them to structured document intermediate
representations. These intermediate representations reflect, but
simplify, the structure of the original documents.  We then build a
model based on pairwise relationships between subparts of each
document, and rank each source file accordingly.  In
\secref{representation} we formalize a general document representation
and then explain how we compare various sub-representations in
\secref{similarity}.  Finally, we formalize the overall technique in
\secref{our-technique}.


\subsection{Structured Document Representation}
\label{sec-representation} 

Both defect reports and source files are represented as distinct
structured documents. In this paper, we use \emph{structured document}
to denote a set of $\langle$name, value$\rangle$ pairs (called ``features''),
where values are well-typed and drawn from non-overlapping parts of
the original artifact.  The types considered include number, string, 
list of strings, and term frequency vector. A
\emph{term frequency vector} is a mapping from terms (i.e., words) to
the frequencies with which they appear in a given text. Term frequency
vectors are often used in natural language processing; we use them to
represent unbounded freeform text such as defect report descriptions
or source code comments. 

We map defect reports to our intermediate form directly. Defect reports
are structured natural language files containing multiple parts, such
as title, description, optional stack trace, project versions
affected, and operating systems affected~\cite{bugquality}. We first
focus on the natural language title and description. We break the text
into a list of terms by splitting on whitespace and punctuation and
converting each term to all lowercase characters; we then construct
term frequency vectors from the resulting multiset of words. 
Additionally, we also parse and record categorical data, such as the
operating system and software version, representing them as discrete
values in the structured document (e.g., as strings and numbers).
Finally, we parse any stack traces into ordered sequences of strings. 

Source code, which is not expressly written in natural language, is
handled similarly, but with a few extensions that have been shown to
be effective in previous work involving textual
analysis~\cite{shepherd07,yoshida10}.  
We obtain an initial list of terms by
splitting on whitespace and punctuation. However, we obtain further refined
terms by taking advantage of paradigms such as Hungarian notation,
camel case capitalization, and the use of underscores to separate
terms in a single string~\cite{shepherd07}.  For example, given the string
``nextAvailableToken'' we increment frequencies for the following
terms: ``next'', ``available'', ``token'', and ``nextAvailableToken''.
Source files are also structured and can also be decomposed into
substructures. Substructures include method signatures, method bodies,
comments, and string literals, among others. In addition to the
overall term frequency vector for the entire file, each substructure
is processed separately into its own term frequency vector. Thus our
intermediate representation for a source file will include a term
frequency vector for words in comments, one for words in method
bodies, and so on. 

Finally, many mature software projects also maintain revision
information.  We utilize two specific forms of version control
information: human-written change log messages and frequency of
revision (``code churn'').  When a developer makes changes to one or
more files, the common practice is to include an informative message
when updating a central repository. These messages often explain
both ``what'' the change does (e.g., ``add bounds checks when
receiving socket data'') and ``why'' it was made (e.g., ``fix
high-priority buffer overrun in networking code'')~
\cite{buse10,zimmermannChurn08}. 
These messages thus relate developer concepts and vocabulary to 
particular files, and can thus be used to aid fault localization. The
logs are parsed as basic natural language and included as a
substructure of all relevant source files.  Additionally, we record
the dates at which a file is changed over the lifetime of the project
as a source code feature to compare with the submission
date of the defect report.  This feature may be helpful in fault
localization as previous work has shown that historical ``code churn''
is often a good predictor of which files will be changed in the
future~\cite{ball_churn05}.



\begin{figure}[t]
\begin{center} 
\scalebox{0.75}{\includegraphics[width=4.4in]{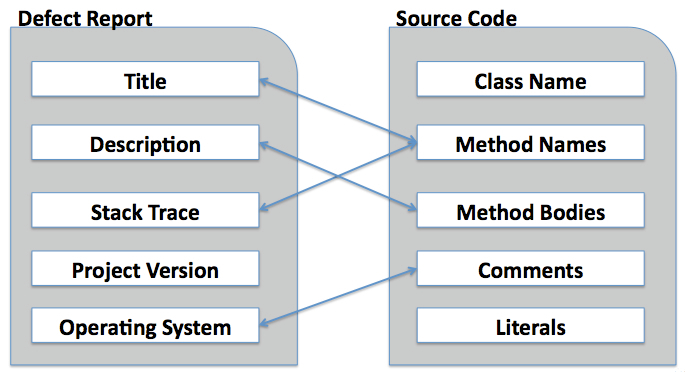}}
\caption{Architecture for fault localization via natural language.
Defect reports and source code are mapped down to structured
documents. The substructures can then be compared pairwise (not all shown) 
using 
separate metrics (e.g., the report title text might be compared to
the source file method names using a term frequency vector comparison,
while the report stack trace might be compared to the source file
method names using a positional index). The overall similarity, and
thus the fault localization rank of that source file, is the weighted
sum of the substructure similarities. 
\label{fig-structured-doc}
\vspace{-0.3cm} 
} 
\end{center}
\end{figure}

Once we have reduced the source code and the defect report to
structured documents, we can compare their substructures pairwise  
to determine their similarity. \figref{structured-doc} shows
an example of this overall approach, with only some of the pairwise 
comparisons highlighted. Next, \secref{similarity}
describes our approach to comparing term frequency vectors, and
\secref{our-technique} shows our overall model for fault localization.

\subsection{Textual Document Similarity}
\label{sec-similarity}

Intuitively, two documents are similar when their subparts have a large
fraction of their terms in common. The more terms the two corresponding
pieces of text share, we assume, the more related concepts they both
describe.

In practice, some terms are more indicative of underlying similarity
than others. For example, terms such as ``int'', ``class'' or ``the'' 
may occur frequently in two unrelated documents. We wish to limit the
impact of such terms on our similarity metric. However, since we
desire a language-independent approach, rather than hand-crafting an
\emph{a priori} stop-list of common words to discount, we derive
that information from the set of available defect reports and
source code. Intuitively, two documents that share a rarer term, such
as ``VerticalRuler'', should be measured more similar than two
documents that share a common term such as ``int''. 

To formalize this intuition we use the \emph{term frequency --- inverse
document frequency} (\textsf{TF-IDF}) measure~\cite{jones72}, which is
common in information retrieval tasks.  We want to measure how strongly
any given term describes a document with respect to a set of context
documents.  Given a document $d$ and a term $t$, the \textsf{TF-IDF}
weighting 
is high if $t$ occurs rarely in other
documents, but relatively frequently in $d$. Conversely, a low weight
corresponds to a term that is frequent globally and/or relatively
infrequent in $d$. The weight for a document $d$ and term $t$ is
computed as follows:
%

$$
\begin{array}{c}
\mathsf{tf}(t,d) =  \dfrac{\mathrm{\#~occurrences~of~}t \mathrm{~in~}d}
{\mathrm{size~of~} d} \\ \\
\mathsf{idf}(t) = \dfrac{\# \mathrm{~of~documents}}{\# 
\mathrm{~of~documents~containing~}t} 
\end{array}
$$

Note that $\mathsf{idf}$ is defined with respect to a corpus of
available documents. In our experiments, when comparing against a
given defect report, the corpus is taken to be all terms in all source
files, log messages, and defect reports filed before that report in question.
With the background formalisms thus described, we now explain how we
combine them to aid fault localization.

In our approach, the overall similarity between a defect report and a
source file is built up from the pairwise similarities between their
substructures (e.g., term frequency vectors). We wish to
empirically determine which pairwise comparisons are the most
predictive of fault localization when the two structured documents in
question are compared.  For instance, previous work has shown that
defect report titles are highly significant when searching for duplicate
reports~\cite{weimer08jalbert}; we hypothesize that they may be
similarly significant when attempting to locate defects. 
\secref{our-technique} describes such a weighting.  

\subsection{Our Technique}
\label{sec-our-technique}
We build on a portion of the {\sf TF-IDF} formalism 
to form our overall similarity metric: 
$$
\mathsf{similarity}(v_1,v_2) = \sum_{t \in ~ v_1 \cap v_2} { v_1[t]
\times v_2[t] \times \mathsf{idf}(t) } 
$$
For each term contained in both documents' parts, we multiply the product of
its frequencies in both documents subparts by that term's {\sf idf} weight.  While
we use {\sf idf} in a standard way, we measure term frequency based only on the 
number of occurrences of a given term without normalizing based on document 
size. 
The aggregate sum over all words' values then serves as the similarity
measure for those two documents' subparts. 

A major distinction between this
metric and standard approaches is that we do \emph{not} normalize for
the size of the documents. While normalization is natural in many
information retrieval tasks, we claim that the special structure of
source code and the fault localization task make it undesirable here. 
For example, consider a defect report that mentions the term
``VerticalRuler'' in a project where the only source code reference to
that term occurs inside one very large source file. In such a case, we
would like to report that single source file as very similar to the 
defect report. However, if the file's size were normalized, it would
appear to be less similar to the defect report
than smaller files that share more common terms (e.g., ``database'').  
Additionally, large projects often contain many \emph{code clones}, and
while not all cloning is harmful~\cite{kapser06}, much of it is
inconsistent: for example, code clones are changed consistently a mere
45--55\% of the time~\cite{krinke07}. 
In standard information retrieval, near-duplicate text may be an
uninteresting search result, but when looking for defects, near-duplicate
code clones should all be considered. We want to
account for the possibility of higher concentrations of code clones
in larger files and not discount the associated natural language artifacts
based on file size alone.  In general, other works apply size-based
normalization when large documents increase false positive rates or
otherwise degrade the accuracy of a given method. We claim that the loss of
precision associated with normalization outweighed the benefits it
provided. This is in line with previous claims~\cite{shepherd07} that
traditional information retrieval search techniques used for documents do
not map perfectly to code-based textual analysis. 

While the above technique is intended for use with two term frequency
vectors, we require certain adaptations for other types of structured
data. Categorical data, such as operating system flavors or program
versions, are treated as a vector with a single term and the metric
can be used in the standard fashion.  Stack trace vectors ---
sequences of strings representing method names --- are compared as word
vectors by using the inverse positional index of a method in the call trace
name as its frequency (thus weighting the first method the highest). 

Finally, we incorporate the idea of code ``churn'' into the technique as it
has shown to correlate with defect density and is readily measurable given
a source repository~\cite{ball_churn05}. Formally, we measure a source
file's degree of ``churn'' by counting the number of times the file was
changed during a set window of time. In our experiments, we used the
entire available source history as the time window, equating code churn
with the number of changes that had been checked in against a file. Similar
to categorical data, we treat ``churn'' as a vector with a single term.       

Given a defect report $D$ and set of source files $f_1 \dots f_n$, 
our goal is to produce a rank-ordered list of the files, weighted
such that files likely to contain the defect are at the top. Human
developers then inspect the files on the ranked list in order until
the fault has been localized. The rank of a file $f_i$ is given as
follows: 
\[
\mathsf{rank}(D,f_i) = \sum_{v_j \in D} \sum_{v_k \in f_i} c_{jk} \mathsf{similarity}
(v_j,v_k)
\]
where $v_j$ ranges over all of the term frequency vectors in the
defect report's intermediate representation, $v_k$ ranges over 
all of the term frequency vectors in the source file's intermediate
representation, and each $c_{jk}$ is a weighting constant for that
particular vector pair. The $c_{jk}$ constants are the formal 
model: a high value indicates that similarity in the associated pair
of sub-substructures (e.g., defect report title paired with source
code comments) is relevant to fault localization. 

One approach would be to use machine learning or regression to
determine the values for the $c_{jk}$ weightings. The size of our
dataset, which includes tens of millions of datapoints and all terms
in over 48,000 files and 6.5 million lines of code, precludes such a
direct approach, however. Attempts to apply linear regression to the
dataset failed to terminate on a 36 GB, 64-bit eight-core 3.6 GHZ machine
within four hours. For scalability, we instead use several common
statistics as a starting point for a parameter space optimization to obtain
a model (see \secref{training-model}). 


\section{Evaluation}
\label{sec-eval}

We conducted two main experiments to evaluate our approach. The first
directly compares the accuracy of our technique to other lightweight
baselines at file-level localization and indirectly compares to
state-of-the-art techniques.  The second experiment quantitatively
verifies our hypothesis that fault localization via textual analysis
depends significantly on human word choice. 

\subsection{Subject applications and defects}

\label{sec-subjApps}
The experiments used three large, mature open source
programs and 5345 total defect reports, shown in \figref{subject_apps}.

We chose these projects for several reasons. First, they are
relatively indicative of substantial, long-term real-world development in
terms of size (at least 6.5 million lines of code total) and maturity (each
is 8 to 11 years old). Additionally, each project has both defect report
and source code repositories.

For each program, we obtained the subset of the available defect reports
for which we could establish a definitive link between the report
and a corresponding set of changes to source files.  We thus
restricted attention to those defect reports that were mentioned by
number in source control log messages. We additionally restricted
attention to reports of actual faults, omitting feature requests and
other invalid or duplicate reports filed using the defect report system.
Also, we only considered defects for which all corresponding changes
took place in source files (i.e., \t{.java}, \t{.cpp}, etc., but \emph{not}
\t{.xml}) in the main
branch of each project (e.g.,
omitting changes to minor branches, testing branches, or data
files).\footnote{Eclipse's \t{/cvsroot/eclipse}; Mozilla's \t{/cvsroot}; OpenOffice's \t{/
trunk}. }  
Accordingly, the numbers for ``files used'' and ``code used'' in
\figref{subject_apps} correspond to source files in trunk of each project's
repository. Finally, we excluded files or reports that could not be
processed (e.g., from CVS or parsing errors). 

To avoid over-fitting the model to our chosen data set, we split the
defects into separate training and testing subsets.  To train the
model, we selected 450 (8\%) of the defects that occurred first
chronologically across all three projects.  The model was created and
refined using only these preliminary defects and the remaining 4915 defects
were held out to evaluate the model. This suggests that such an approach
could be implemented by collecting a minimal number of initial defects
while achieving high accuracy as reported in the following
sections.\footnote{We could have used
cross-validation~\cite{kohavi} instead to help detect bias from
over-fitting, but prefer to use holdout validation because of the large
number of available datapoints and the time-series nature of the data:
defect reports often make reference to
previous defect reports~\cite{bugquality,weimer08jalbert}. It would not be
valid to train a model on future defect reports and evaluate it on past
ones.} 


\begin{figure*}[t]
\begin{center}
\begin{tabular}{|l|r|r|r|}
\hline
Project (Date checked out) & Source files with defects & Total source files & Percentage\\
\hline
Eclipse (2009.09.08) & 2660	 & 21303 & 12.49\% \\
\hline
Mozilla (2009.09.08) & 1811	& 6179 & 29.31\% \\
\hline
Openoffice (2009.09.22) & 1463 & 1507 & 97.08\% \\
\hline \hline
Total	 & 5934 & 28989 & 20.47\% \\
\hline
\end{tabular}
\end{center}
\caption{Distribution of files with defects throughout the subject
applications.  In this case, a file is said to contain a defect if it
was changed in a repository revision that specifically mentioned
fixing a certain defect, and addressing that defect involved only
changes to source files.\label{fig-defect-dist}
}
\end{figure*}

\begin{figure*}[t]
\begin{center}
\begin{tabular}{|l|r|r|r|r|r|r|}
\hline
Program & Total Defects & Files & Lines of & Language(s) & Avg. report    & Avg. 
report \\
        & Used    & Used   & Code Used&             & length (lines) & title (words)  \\
\hline
\hline
Eclipse & 1,272 & 23,601 & 3,476,794 & Java & 172.535 & 8.642\\
\hline
Mozilla & 3,033 & 14,651 & 2,262,877 & Java, C++ & 316.811 & 9.428\\
\hline
OpenOffice & 1,040 & 9,992 & 815,473 & Java, C++ & 60.547 & 5.623\\
\hline
\hline
Total & 5,365 & ~~48,244 & 6,555,144 & - & - & - \\
\hline
\end{tabular}
\end{center}
\caption{Subject programs used in our evaluations. ``Defects'' counts
reports that could be linked to a
particular set of changes. ``Files'' counts retrieved source files
in the project branch, including those not involved in defect reports.
``Lines of Code'' measures the size of those source files, while
``Languages'' lists their programming languages. The last two columns
measure aspects of the defect reports used. 
\label{fig-subject_apps}
\vspace{-0.3cm} 
}
\end{figure*}

\subsection{Parameter space optimization}
\label{sec-training-model}

Our first step is to build a model relating similarity comparisons
between defect report and source code structures to fault
localization. In the terminology of \secref{our-technique}, this
involves determining values for the 34 distinct $c_{jk}$ weights. 

To build such a model we first performed a one-way analysis of variance
(ANOVA) on a subset of the data to estimate the predictive power of
each possible document comparison. For each defect report we consider
all of the files that were eventually fixed by the developers and also
150 files, chosen at random, that were not.\footnote{The inclusion of
150 files was chosen to be as large as possible while allowing the
problem to be tractable on available hardware; see
\secref{our-technique}.} We pair each such file
$f_i$ with the original defect report $D$ to produce one datapoint.
Each datapoint has multiple associated features (i.e., the explanatory
variables): there is one feature for each each of the 34 $\langle v_j,
v_k \rangle$ vector pairs, with the measured similarity serving as the
feature value. The response variable for a given datapoint is set to
$1$ if the file was modified by developers and $0$ otherwise. 

An ANOVA measures the ratio of the variance explained
by each model feature (i.e., each $\langle v_j, v_k \rangle$ similarity)
over the variance not explained.  
We use this ratio only as a starting point for $c_{jk}$; distant
values will merely yield a longer training search time. These ANOVA values
may not be optimal because our final model goal is to rank order the files
for fault localization and not to minimize the error between a model and
the artificial $0$ and $1$ response variables.  

The second step was to perform a principle component analysis (PCA) to
determine the number of components that were relevant to the task of
detecting the location of a fault in source code. Given our 34 possible
document substructure comparisons, this analysis showed that a
combination of 15 accounted for more than 99\% of the overall variance
in the data. 
The final $c_{jk}$ values were obtained via a gradient ascent parameter
space optimization. In each iteration, the best model available
was compared to similar models, each constructed by increasing
or decreasing the value of a single $c_{jk}$ by 10\%. The comparison
was conducted using the \emph{score} metric detailed in
\secref{metric}. We terminated
the process when the improvement between one iteration and the next
was less than 0.01\%; this took 5 iterations. We used
the final $c_{jk}$ values as our formal model.  


\subsection{Experiment 1 --- Ability to localize faults}
\label{sec-secExp1}
Our first experiment measures the accuracy of our technique when 
localizing faults. We compare two versions of our
technique against two baseline approaches directly. We also
indirectly compare against the published results of three state-of-the-art
tools using a common metric.  

\label{sec-metric}
We adopt the \emph{score} metric for measuring the accuracy of a fault
localization technique. The \emph{score} metric is commonly used in fault
localization research~\cite{cleve05,harrold05,Renieris03}.  
As described in previous work, ``the score defines the percentage of the
program that need not be examined to find a faulty statement in the
program.''~\cite[p.~6]{harrold05} For example, a ranking for an
OpenOffice defect report that requires the user to inspect 2,000 of
the 9992 files before finding the right file has a \emph{score} of
$(9992-2000)/9992 = 80\%$. Higher \emph{score} values indicate better
accuracy. We apply the \emph{score} metric at the file level of granularity.
We report the average \emph{score} over all defects available. 

%

\begin{figure*}[t]
\begin{center}
\begin{tabular}{|l|r|r|r|r|r|}
\hline
Test Set & \# Defects & Our Approach & Stack Trace & Code 
Churn & Optimal Search \\
\hline
\hline
OpenOffice only & 1018 & 82.728\% & 57.979\% & 72.755\% & 75.731\% \\
\hline
Eclipse only & 1124 & 89.937\% & 56.295\% & 73.131\% & 91.155\% \\ 
\hline
Mozilla only & 2773 & 95.359\%& 50.152\% & 93.860\%  & 87.906\% \\
\hline
Stack traces only & 325 & 89.608\% & 65.060\% & 76.442\%  & 90.683\% \\
\hline
\hline
\textbf{Complete set} & {\bf 4915} & \textbf{91.502\%} & \textbf{53.137\%} & 
\textbf{84.820\%}  & \textbf{86.128\%} \\
\hline
\end{tabular}
\vspace{-.3cm} 
\end{center}
\caption{\emph{Score} values for selected techniques. The ``Test Set''
column lists examined subsets of the 4915 defects from three programs;
450 separate defects were used as training in 
Section \ref{sec-training-model}.
``Our Approach'' measures the \emph{score} obtained by our technique. 
The ``Stack trace'' baseline favors files mentioned in user-provided
stack traces, the ``Code churn'' baseline favors
frequently-changed files, and the ``Optimal search'' baseline simulates
an optimal code search based on defect report terms. \label{fig-exp1}} 
\end{figure*}

\figref{exp1} shows the results. A lower baseline of 50\% represents
inspecting files in random order. Our approach
outperforms all baselines over the entire test set (highlighted
in boldface in \figref{exp1}) and is generally better than
other approaches in most subsets.  The ``Stack traces only'' subset
includes all defect reports that featured stack traces. Note
that of these 4915 defect reports used to evaluate our approach, 
only 325 (6.7\%) contained stack traces. 

We compare against three baselines motivated in
\secref{motivation} by mimicking some steps developers might
take when attempting to fix a defect.  In each case we produce a
ranked list and compute a score metric to admit a direct comparison with our
technique. The \emph{code churn baseline} ranks files in descending
order of number of changes throughout the entire history of the
system up to the date of the defect report in question. The
\emph{stack trace baseline} ranks files by their position in
any stack trace provided as part of the defect report; all files not
mentioned in the stack trace are equally likely to be chosen after
all files mentioned in the trace. Finally, the \emph{optimal search
baseline} approximates a developer using a search tool with 
some degree of domain knowledge. Given a search term,
such a tool can return a list of all source files mentioning
that term, ignoring case, ranked by number of occurrences. All files
that do not contain the term are equally likely to be chosen after
any files that do contain the term. The optimal search baseline
considers every word in the defect report and uses the one that
yields the best score result (i.e., the search term that indicates
all of the relevant files and as few irrelevant files as possible). 
Note that the best search term cannot, in general, be known \emph{a priori}
by an automated technique: using only the best term is meant to represent 
human knowledge of the software system. While this baseline does not
perfectly model human code search, it approximates
the process for the purposes of comparison. 

Over the entire test set, we outperform the stack trace, code churn
and optimal search baselines by 38\%, 7\%, and 5\% respectively.
While the performance gain over a stack trace baseline is immediate,
the lower performance gain over code churn overall and over optimal
search within the Eclipse and Mozilla projects requires more of an explanation.
Eclipse has the most source files out of the three benchmarks.
Additionally, as noted in \figref{defect-dist}, the bugs are localized
to only 12.49\% of the files.  Both baselines thus have the potential
to eliminate much of the search space.  However, the results for both
baselines for OpenOffice show that this
is not the general case. In addition, since both external baselines provide
overall \emph{scores} of over 84\% and 86\% respectively, only a 
16-point and 14-point \emph{score} increase is
possible in either case.  In that regard, our 7-point and 5-point increases 
constitute nearly 44\% and 36\% of the respective remaining room for 
improvement. Finally, on large projects, even
small gains are significant: for example, the 7\% \emph{score} increase over
the code churn baseline prevents an aggregate
of 4,501 source files (or 611,812 lines of code) from being considered
during the fault localization search over our entire test set.


Our technique performed most poorly on OpenOffice defects: if only
Eclipse and Mozilla are considered, our performance is almost 94\%. This
is explained by a quirk of the OpenOffice project:
their defect reports are about four times smaller (see
\figref{subject_apps}), thus reducing one of the primary sources of textual
similarity (see \secref{representation}). 

The results presented in \figref{exp1} show that our tool outperforms
lightweight baselines. We also suggest that our technique
may perform better than more heavyweight techniques. 
Several state-of-the-art fault localization techniques report accuracy
values for their tools in terms of the distribution of subject faults
over the scale of possible \emph{score} measures. For comparison
purposes we use a weighted average of each \emph{score} interval to
calculate an overall accuracy measure for each approach.  The tools of
Jones \textit{et al.}~\cite{harrold05}, Cleve \textit{et
al.}~\cite{cleve05}, and Renieris \textit{et al.}~\cite{Renieris03}
achieved aggregate \emph{score} measures of 77.797\%, 63.415\%, and
56\% respectively. The largest of these projects evaluated on 132
defects over seven files containing at most 560 lines of code each. 
While these results are measured on different test sets and are
therefore not directly comparable, we note that our technique
obtains a \emph{score} result 14 points higher than previous work and
is evaluated on an order-of-magnitude more defects and files. 

Finally, our technique is lightweight in terms of execution time.
Assuming code files are kept indexed as word vectors, our tool always
runs in under 10 seconds per defect report and generally takes less
than 1 second.  

\subsection{Experiment 2 --- Human word choice}

Our second experiment tests our hypothesis that our \emph{score}
accuracy is mainly due to correctly extracting and comparing the
natural language chosen by humans in defect reports and source files.
We first demonstrate that our technique's accuracy is not dominated
by other features, such as length, defect priority, or defect lifespan. 
Secondly, we alter the natural language of
the subject reports systematically, showing that performance degrades
in a proportional manner. Finally, we evaluate the relative predictive
power of our model's features. 

\begin{figure}[t] 
\begin{center}
\begin{tabular}{|l|r|}
\hline
Document feature & Correlation with \emph{score} \\
\hline
\hline
Average report length & 0.24\\
\hline
Maximum report length & 0.22 \\
\hline
Defect lifespan &  0.20\\
\hline
Rate of commenting in edited source ~ & 0.18\\
\hline
Number of duplicate reports &  0.17	\\	
\hline
Report readability &  0.10\\
\hline
Number of edited source files & 0.08\\
\hline
Reported priority &  0.07\\
\hline
\end{tabular}
\end{center}
\caption{Pearson correlation between surface features and our
technique's \emph{score}. 
\vspace{-0.3cm} 
\label{fig-exp2}
} 
\end{figure}

We hypothesize that human-chosen natural language in defect reports
and source code is a critical factor in our fault localization
approach. We first discount several other potentially-prominent 
features  in terms of predictive power with respect to the
\emph{score} accuracy of our technique.  The features examined
cover both defect reports and source code: the Flesch-Kincaid
readability level of the report in question~\cite{flesch}, the assigned
defect priority, the number of total reports for a defect when considering
all duplicates, the maximum report length for a defect, the average
report length for a defect, the overall lifespan of the defect from reported
defect to reported patch, the number of source files edited as part of
the patch, and the rate of commenting in the edited source code.  

We calculated the Pearson correlation of all 4915 total defects'
\emph{score} measures with these features.  The correlations can be
found in \figref{exp2}.  It is generally accepted that correlations
below 0.3 are not statistically significant~\cite{giventer}. 
All observed correlations fell well within these bounds and therefore
we conclude that these features do not significantly affect our model.
However, of all correlations, report length and rate of commenting had some of
the highest relative values. This supports our claim
that natural language is key to our technique's success, since these
features typically relate directly to the natural
language present.

\begin{figure}[t]
\begin{center} 
\scalebox{0.75}{\includegraphics[height=3in]{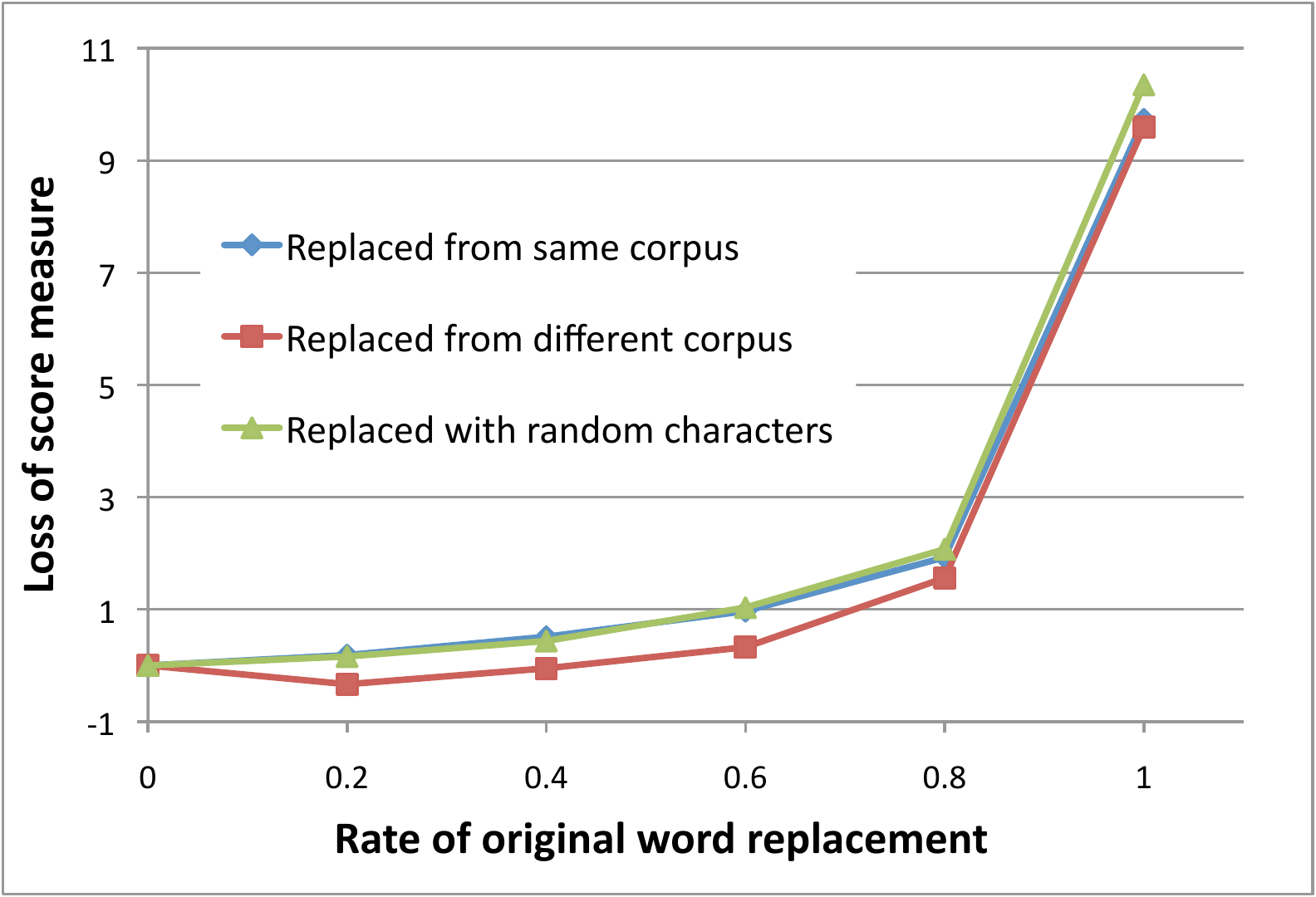}}
\caption{The effect of replacing human-chosen words with various
random words on our technique's
\emph{score} over all 4915 defects in the test set.
\label{fig-fexp3}
\vspace{-0.3cm} 
} 
\end{center}
\end{figure}

Next, we demonstrate that our model is greatly affected by the
users' choice of language in defect reports and the developers' choice
of language in source code. To evaluate this, we measure our
\emph{score} accuracy as more and more human-chosen words are replaced
by random words. 
We used three different random techniques to replace
human-chose words: replacing terms with words
from the same general set (e.g., the set of all report description
words), replacing terms with words from a
different set (in this case, an English dictionary), and finally,
replacing terms with strings of the same length made up of randomly
selected characters (i.e., random noise).  
In each case, we
altered the natural language in increments until the entire frequency
vector had been changed, using the unaltered reports as a baseline.
The results of this experiment can be found in \figref{fexp3}. Each
datapoint represents the degradation in \emph{score} of our algorithm running 
on the
entire 4915-defect testing data subset with some fraction of each defect report's
text altered. 


As the natural language in defect reports is changed, and thus the useful
information in the report is reduced, the performance of our technique
degrades.  The reduction in \emph{score} is not strictly proportional, as
is expected from the presence of common words and our use of the {\sf idf}
weighting: retaining even a few words that account for some of the relevant
document similarity in a given comparison degrades the performance of the
tool only slightly.  In addition, when replacing terms with words from a
different corpus the performance initially increases very slightly and then
decreases, following the other two replacement techniques.  
The sharp increase of degradation when all terms have
been altered further reinforces the idea that our approach can perform
accurately with even a small amount of natural language information and
fails only when almost all information is changed.
The general trend in
\figref{fexp3} is that performance of our approach degrades when natural
language information is removed or altered.  Thus, we posit that our
approach is leveraging the human-chosen language and not additional
features.


\begin{figure}[t]
\begin{center}
\begin{tabular}{|l|l|r|}
\hline
Report Substructure ~ & Code  Substructure ~ & ~ Singleton \emph{score} \\
\hline
\hline
Report body & Method bodies & 84.94 \\
Report date & Code churn & 84.82 \\
Report body & Log message & 84.51 \\
Report body & Comments & 83.60 \\
Component & Log message & 71.62 \\
Report title & Method bodies & 70.44 \\
Report body & Method signatures & 69.39 \\
Report title & Comments & 67.52 \\
Operating system & Comments & 67.31 \\
Component & Method bodies & 65.85 \\
Component & Comments & 64.76 \\
Report title & String literals & 60.69 \\
Report body & Class name & 57.10 \\
Stack trace & Class name & 53.14 \\
Report title & Method signatures & 52.59 \\
\hline
\end{tabular}
\vspace{-.3cm} 
\end{center}
\caption{The score results from a ``singleton'' analysis in which we
used only one feature and measured the score it achieves alone.  
\label{fig-features}
\vspace{-.3cm} 
}
\end{figure}

Finally, having established that human word choice is critical, we 
evaluate which words and comparisons are the most important. 
\figref{features} shows the predictive power of the features in terms
of a ``leave-one-in'' or ``singleton'' analysis.\footnote{The size of
the dataset precludes a full ANOVA; see \secref{our-technique}. In
addition, heavy feature overlap precludes the use of a
``leave-one-out'' analysis; see PCA in \secref{training-model}.} To obtain these
results we built many different versions of the model, each utilizing
only one of the features.  Thus, the score measure for each version
suggests the relative utility of the given feature with respect to
localizing faults. 

The report title and body, as well as the method bodies
and comments, are involved in many of the most useful relationships
in our fault localization model.  With respect to defect reports, the
titles and bodies contain the majority of the natural language
information chosen by the reporter and, as such, are more helpful than
extraneous categorical data and stack traces. Comparatively, we
believe that code revision log messages are helpful because they often
address specific defects or defect reports and thus might use similar language.
Code churn also proves to be predictive of defect location, supporting
claims made in previous work~\cite{ball_churn05}.
Intuitively, code comments are effective when paired against terms from defect
reports because they are both written explicitly in natural language and
often encapsulate code specifications in a manner complementary to the
language inherent in the code's identifiers.  Method bodies contain
most of the text associated within code files and thus also serve as
effective predictors. Finally, more obscure categorical information (e.g.,
processor architecture) and string literals found in code were less useful
to the model.

\subsection{Threats to validity}

Although our experiments are designed to demonstrate that our
technique performs well over a large number of defects and files, our
results may not generalize to industrial practice. First, our
benchmark programs may not be indicative. The programs we chose are
all large, mature, open-source projects. While they span three
individual domains, they may not generalize to all potential domains.
Our results may not apply to younger, smaller projects, but we claim
that fault localization becomes less interesting as the project
shrinks (e.g., in the limit, fault localization is rarely a primary concern
for a project with only a few small source files). In addition, all
of our benchmarks involve GUI components, making them more likely 
to support our hypothesis that report-writers and developers will use
similar textual terms. We view our evaluation on large datasets (e.g., ten
times larger than previously-published
evaluations~\cite{cleve05,harrold05,Renieris03}) as an advantage. 
 
Bird \emph{et al.} note that sampling defect reports for the purpose of
experimentation may lead to biased results~\cite{bachmann2010mlb,bird09}. As 
a result,
our technique may only be good at localizing certain types of faults
(i.e., those that open source developers deign to mention in version
control logs). Lacking a project with a linked
version control and defect repository, we cannot mitigate this threat
beyond our claim that manual inspection of the reports found the faults to
be a relatively even cross-section of each project's repository
over the history of that project (see \secref{subjApps}). 

Our code churn baseline may not be indicative because it relies on
eight to ten years of version control information. For example, it
may perform particularly well on the larger and older Mozilla project,
correctly giving low rankings to the many files that have been stable
for years. In practice, a development organization may not have 
such rich version history information, or such stable files may be
manually excluded by developers. 

Not all files in a project will be associated with fixed defects when using
our selection methodology from \secref{subjApps}. Previous researchers
have noted that in practice, defects are not uniformly
distributed~\cite{ball_churn05}. If our model somehow learned the
underlying defect distribution rather than using information from the defect
reports, it would not generalize. We guard against this threat both by
construction (i.e., our learned features are all coefficients for
similarity metrics \emph{between} defect report substructures and software
development artifacts, and not development artifacts alone) and also
by our use of holdout validation. 

Finally, when comparing our results with those of established fault
localization techniques using the \emph{score} metric, we interpolated
to convert previous \emph{distribution}-style results to single
\emph{score} numbers and thus admit more illustrative comparisons. Previous
publications have reported \emph{score} value \emph{distributions} over
intervals from 0\% to 100\%.  We estimated based on a weighted average of
the medians of each interval. As a result, these indirect comparisons with
previously-published results cannot used to draw firm conclusions and
serve instead to provide descriptive context. 


\section{Related Work}
\label{sec-related}

Related research to our work falls into two main categories: prior
work in fault localization, and prior work in reverse engineering. 

\subsection{Fault Localization}
Ashok \textit{et al.} propose a similar natural language search technique
in which users can match an incoming report to previous reports,
programmers and source code~\cite{Ashok09}.  By comparison, our technique
is more lightweight and focuses only on searching the code and the
defect report.  

Jones \textit{et al.} developed Tarantula, a technique that performs
fault localization based on the insight that statements executed often
during failed test cases likely account for potential fault
locations~\cite{harrold05}.  Similarly, Renieris and Rice use a ``nearest 
neighbor'' technique in their
Whither tool to identify faults based on exposing differences in
faulty and non-faulty runs that take very similar executions
paths~\cite{Renieris03}. These approaches are quite effective when
a rich, indicative test suite is available and can be run as part of
the fault localization process. They thus requires the fault-inducing
input but not any natural language defect report. By contrast,
our approach is lightweight, does not require an indicative test
suite or fault-inducing input, but does require a natural language
defect report. Both approaches will yield comparable performance, and
could even be used in tandem. 

Cleve and Zeller localize faults by finding differences between
correct and failing program execution states, limiting the scope of
their search to only variables and values of interest to the fault in
question~\cite{cleve05}. Notably, they focus on those variable and
values that are relevant to the failure and to those program execution
points where transitions occur and those variables become causes of
failure. Their approach is in a strong sense finer-grained than ours:
while nothing prevents our technique from being applied at the level
of methods instead of files, their technique can give very precise
information such as ``the transition to failure happened when $x$
became 2.'' Our approach is lighter-weight and does not require
that the program be run, but it does require defect reports. 

More recent work conducted by Wang \emph{et al.} aims to refine 
the concept of fault localization based on test suite coverage 
metrics~\cite{wang09}.  They closely examine contextual information to detect
faults that are being executed but not identified.   

Liblit \textit{et al.} use Cooperative Bug Isolation, a statistical
approach to isolate multiple defects within a program given a deployed user
base. By analyzing large amounts of collected execution data from real
users, they can successfully differentiate between different causes of
faults in failing software~\cite{liblit05}. Their technique produces
a ranked list of very specific fault localizations (e.g., ``the fault
occurs when $i > arrayLen$ on line 57''). In general, their technique
can produce more precise results than ours, but it requires a set of
deployed users and works best on those defects experienced by many users.
By contrast, we do not require that the program be runnable, much less
deployed, and use only natural language defect report text. 

Jalbert \emph{et al.}~\cite{weimer08jalbert} and Runeson \emph{et
al.}~\cite{Runeson07} have successfully detected duplicate defect reports
by utilizing natural language processing techniques. We share with
these techniques a common natural language architecture (e.g.,
frequency vectors, {\sf TF-IDF}, etc.). We differ from these
approaches by adapting the overall idea of document similarity to work
across document formats (i.e., both structured defect reports and also
program source code) and by tackling fault localization. 

\subsection{Reverse Engineering}
Latent Semantic Indexing (LSI) is an information retrieval
technique for measuring document similarity~\cite{Deerwester90}.
Similar to our technique, it uses word frequency vectors to measure 
co-occurrence of relevant terms in documents.  Marcus and Maletic
used LSI to expose document-to-source-code 
traceability~\cite{marcus03}.  While their work mainly focuses on
matching documents from the initial phases of the development process,
the work presented in this paper attempts to match specifically defect
reports created by both users and developers throughout the
maintenance process.  Additionally, traditional LSI treats documents
as a single, unified term frequency vector whereas our technique breaks
documents down into substructures based on the hypothesis that certain
language is more helpful for localizing defects. 

Li \emph{et al.} have examined the problem of extracting information
from structured documents in addition to categorizing that 
information~\cite{li09}. They focus on user queries in particular,
which is similar to the defect reports we study in this work.  They
also note that tailoring analyses to specific corpora is particularly
helpful, which we confirm with the use of inverse document frequency
for weighting individual terms.

Devanbu \emph{et al.} and W\"{u}rsch \emph{et al.} both developed 
software system search tools that leverage the natural language in both
source code and related software artifacts~\cite{devanbu90,wursch10}. 
While our system has a similar
back-end natural-language-based approach, our overall goal is automatic 
fault localization, not general code search or program comprehension.

Breu \emph{et al.} studied the effect of user interaction throughout the 
defect fixing process~\cite{breu10}.  Much like the work presented in this
paper, they found that additional information and user clarification generally 
only serves to aid in fixing defects.  While their focus is on the
interaction between users and developers throughout the maintenance
process, our work aims to measure the quality of information contained in
different parts of documents associated with defect fixing. 

Ko \emph{et al.} have studied the overall process used by developers 
to find information and understand programs~\cite{Ko06}.
They employed a human study to gain insight into what kinds of 
information developers think is relevant to a given task and how they 
make decisions in this process.  The goal of our research is complimentary 
to this work in that we are trying to automatically discern which information
is related to specific defects and, in a broader sense, aid in the maintenance
process on the whole.

Shepherd \emph{et al.} focused on both proving that the natural
language in source code is meaningful and also on attempting to
extract those language artifacts in a meaningful and useful
manner~\cite{shepherd07}.  They studied natural language use in code
for the purpose of developing a specialized code-search technique
specifically focused on identifying distributed concepts throughout a
system.  Similarly, Lawrie \emph{et al.} have examined the quality of
source code identifiers in terms of code
comprehension~\cite{lawrie07}.  They show that insightful and
carefully chosen natural language identifiers make for more
understandable and maintainable code.  We build upon such work by
leveraging these facts in the domain of fault localization. 

Much work has been done to measure the quality of natural language choices
made by developers~\cite{Caprile00,Deissenboeck06,lawrie07,Takang96}.
Additionally, some of this work looks at restructuring or refactoring
natural language artifacts in an attempt to reverse engineer the original
developers' intentions and aid program understanding.  We claim that
measuring the quality of natural language is orthogonal to the work we
present in this paper. We are more concerned with the ability of the
natural language in both defect reports and source code to localize faults,
regardless of the language's \emph{quality}.  While higher quality
information may allow our tool to compare documents more accurately, our
tool currently achieves very high accuracy 
without accounting for the quality of the underlying natural language.


\section{Conclusion}
\label{sec-conclusion}
\label{sec-future}

We present a lightweight, scalable technique for localizing faults
based on document similarities.  We hypothesize that human-chosen
natural language present in both defect reports and source code can
be compared to identify potential fault locations based
on natural-language descriptions. Our technique is entirely static and
is language independent. 

An empirical evaluation shows that our technique not only performs
better than several baseline approaches, but is comparable to the
state-of-the-art techniques without requiring significant overhead or
a runnable program and a test suite. We also demonstrated that 
the word choice in natural language artifacts was truly the
dominant factor in our approach.  

A large empirical evaluation of our program on 5345 historical defects
from three real-world programs totaling 6.5 million lines of code
showed that we can reduce the search space for finding a fault by
over 91\% on average. We believe that this approach has the potential
to significantly decrease the cost of fault localization, and thus
software maintenance overall. 


\bibliographystyle{abbrv}
\bibliography{weimer} 

\end{document}